# Aerial hyperspectral imagery and deep neural networks for high-throughput yield phenotyping in wheat

Ali Moghimi[a,1], Ce Yang[a], James A. Anderson[b]

[a] Department of Bioproducts and Biosystems Engineering, University of Minnesota, 1390 Eckles Ave, St. Paul, MN 55108, USA
[b] Department of Agronomy & Plant Genetics, University of Minnesota, 991 Upper Buford Circle, St. Paul, MN 55108, USA
moghi005@umn.edu
ceyang@umn.edu (corresponding author)
ander319@umn.edu

## Abstract

Crop production needs to increase in a sustainable manner to meet the growing global demand for food. To identify crop varieties with high yield potential, plant scientists and breeders evaluate the performance of hundreds of lines in multiple locations over several years. To facilitate the process of selecting advanced varieties, an automated framework was developed in this study. A hyperspectral camera was mounted on an unmanned aerial vehicle to collect aerial imagery with high spatial and spectral resolution. Aerial images were captured in two consecutive growing seasons from three experimental yield fields composed of hundreds experimental plots (1×2.4 meter), each contained a single wheat line. The grain of more than thousand wheat plots was harvested by a combine, weighed, and recorded as the ground truth data. To leverage the high spatial resolution and investigate the yield variation within the plots, images of plots were divided into sub-plots by integrating image processing techniques and spectral mixture analysis with the expert domain knowledge. Afterwards, the sub-plot dataset was divided into train, validation, and test sets using stratified sampling. Subsequent to extracting features from each sub-plot, deep neural networks were trained for yield estimation. The coefficient of determination for predicting the yield of the test dataset at sub-plot scale was 0.79 with root mean square error of 5.90 grams. In addition to providing insights into yield variation at sub-plot scale, the proposed framework can facilitate the process of high-throughput yield phenotyping as a valuable decision support tool. It offers the possibility of (i) remote visual inspection of the plots, (ii) studying the effect of crop density on yield, and (iii) optimizing plot size to investigate more lines in a dedicated field each year.

1. Present address: Department of Biological and Agricultural Engineering, University of California-Davis, One Shields Avenue, Davis, CA 95616, USA (amoghimi@ucdavis.edu)

# 1 Introduction

Considering the increasing world population and subsequent demand for food, crop production should double by 2050 (Tilman et al., 2011), indicating the average rate of yield increase of crops should be 2.4% annually (Ray et al., 2013) – the current average rate of increase is only 1.3% (Araus et al., 2014; Ray et al., 2013). These statistics noticeably indicate an urgent need for further efficiency improvement in crop production to alleviate the global concern of food security. Nevertheless, genetic gain in yield of wheat, one of the major crops, was reported to be less than 1%, far behind the necessary yield increase (i.e., 2.4%) (Crain et al., 2018; Ray et al., 2013; Reynolds et al., 2012). Other studies even claimed wheat yields have plateaued in some regions of the world (Acreche et al., 2008; Araus et al., 2018; Sadras and Lawson, 2011), indicating the importance of high-throughput phenotyping for developing wheat varieties with high yield potential in a more efficient and effective manner.

To identify wheat varieties with high yield potential, plant scientists and breeders examine hundreds to thousands of new candidate lines, developed through breeding and genotyping, in experimental plots each year and measure their yield performance. The yield measurement of wheat plots is performed through conventional methods which rely on demanding, extremely laborious, and time-consuming tasks. For instance, in an experimental yield nursery composed of hundreds of wheat plots, the steps of yield measurement include harvesting the grains of each plot, manual packaging, labeling, and sealing – all steps are repetitively performed for each plot to avoid blending grains of plots. These exhausting tasks escalate even more since breeders have yield nurseries in multiple locations to account for non-uniform climate, soil, and environmental conditions. Furthermore, in a rather short harvesting time, conventional measurement for yield phenotyping is restricted by the availability of machinery, labor and weather conditions. Each of these factors could potentially postpone harvesting time for several days during which yield loss can occur because of animals' attack (e.g. birds and rodents) and/or severe weather (e.g. hail and winds). Any of these challenges could deteriorate the quality and reliability of the data, thus wasting the enormous efforts made thorough the entire growing season.

The other limitation associated with the conventional yield phenotyping methods is that it ignores the spatial variability of yield within the experimental plots. Various regions in a single plot contribute unequally to the measured yield for the plot (i.e., yield is non-uniformly distributed within an experimental plot). Therefore, breeders are unable to study the effect of crop density on yield potential for various varieties. Moreover, ignoring the variability of yield within plots entails an enormous loss of information regarding the marginal effects on yield. This is a valuable information to identify the lines whose plants located in the middle of plot can compete for nutrition and therefore can contribute to yield as much as the plants located at the margin of the plot. Considering the importance of selecting high-yielding varieties and limitations associated with conventional phenotyping methods, there is a compelling need to predict yield, preferably with high-resolution, using robotics equipped with advanced sensing technologies.

In several studies focusing on high-throughput field phenotyping for yield estimation of wheat, researchers have utilized various sensors mounted on unmanned aerial vehicles (UAVs). Madec et al. (2017) attempted to predict the yield of various wheat genotypes based on maximum plant height estimation using RGB images and LiDAR data collected by a UAV. They reported a low correlation between yield and maximum plant height derived from LiDAR data ($R^2 = 0.22$) and RGB images ($R^2 = 0.13$). Duan et al. (2017) computed normalized difference vegetation index (NDVI) derived from multispectral images captured by UAV to predict the yield of wheat in high-throughput phenotyping context. Because of the attained spatial resolution (2-5 cm), the NDVI calculated per each pixel was a combination of vegetation and background, inherently with different spectral characteristics. To address the mixed pixel issue, they proposed a naïve solution in which pixels with NDVI less than a predefined threshold were masked for further analysis. They suggested that there is a high correlation ($R^2 = 0.87$) between the adjusted NDVI, computed around flowering time, and the final yield. However, this finding was achieved from few plots (in total 12 plots including three cultivars with four treatments).

To predict the yield of a particular winter wheat, Du & Noguchi (2017) deployed stepwise regression to analyze five color vegetation indices derived from multi-temporal color images captured by a UAV from heading stage to ripening stage. They performed the analysis on only nine samples of wheat yield. Their results demonstrated a strong correlation ($R^2 = 0.94$ and RMSE = 0.02) between four color vegetation indices and yield for this limited number of samples. In another study, aerial images acquired from UAV were utilized to estimate the yield of twenty wheat varieties under a water limited and heat stressed environment (Kyratzis et al., 2017). Two vegetation indices including green normalized difference vegetation index (GNDVI) and NDVI were calculated at various growing stages of plants over two consecutive years. They concluded that GNDVI, compared to NDVI, performed better in explaining variability of grain yield with $R^2 = 0.31$ and $R^2 = 0.21$ for the first and second year of experiment, respectively.

Nowadays, with the commercialization of UAVs and increasing availability of compact, inexpensive, and sophisticated sensing technologies, the challenge in high-throughput phenotyping shifted from data collection to data analysis – extracting significant features and recognizing underlying patterns from large datasets captured with high temporal, spatial, and spectral resolution by autonomous platforms equipped with non-contact sensing technologies. The common approach for analysis of image-based data (RGB, multi- or hyper-spectral images) is to calculate spectral vegetation indices derived by simple arithmetic equation (e.g. ratio) among a few spectral bands. Nevertheless, more advanced analysis methods are required to extract valuable information from images for high-throughput field phenotyping rather than simple vegetation indices which entail several limitations. For instance, it has been proved that NVDI, the most widely used index, suffers from saturation issue over vegetation canopy with moderate-to-high level of density (Asrar et al., 1984; Gitelson, 2004; Gitelson et al., 1996). Therefore, more advanced analytical approaches are required for high-throughput analysis of large image-based phenotyping datasets.

Recently, machine learning and deep learning algorithms have shown considerable promise in developing more efficient and effective pipelines for analysis of large phenotyping datasets (Singh et al., 2016, 2018). Deep learning (DL), inspired by the biological neural structure in the human brain, refers to computational non-linear models composed of various processing layers in which an abstract representation from the output of the previous layer is learned up to the output layer where a complex function is learned in terms of these abstract representations (LeCun et al., 2015).

Yield is the most fundamental trait in plant breeding since almost every other characteristic of crops, treatments, and management decisions are evaluated through the lens of whether they promote or hinder the yield potential. The primary objective of this study was to develop a sensor-based, automated framework for high-throughput yield phenotyping of wheat in the field. The data from hundreds of wheat varieties were collected by a hyperspectral camera mounted on a UAV flying over three experimental wheat plots during two consecutive growing seasons. To analyze high-dimensional hyperspectral images captured with high spatial and spectral resolution, a deep neural network was trained to predict the yield of wheat plots. In addition to yield prediction at plot scale, the feasibility of yield estimation at a finer spatial resolution (i.e., sub-plot scale) was investigated to determine the ability of wheat lines in producing a uniform yield across the plot – a valuable new factor that can be used in breeding programs to nominate advanced cultivars for commercialization.

# 2 Materials and methods

## 2.1 Field site and experimental setup

Field experiments were conducted in three experimental yield trial fields (C3, C4, and C9) during two consecutive growing seasons 2017 (C3 and C9) and 2018 (C4). Field sites were located at St. Paul Campus Research Facility, University of Minnesota, MN (44°59'28.15"N and 93°10'48.34"W) (Figure 1). Yield trials were composed of hundreds experimental new wheat lines, developed at University of Minnesota, several check lines, and advanced lines from other breeding programs. Each wheat line was planted in seven rows which formed a plot with about one-meter width and 2.4-meter length. The plots were harvested with a combine designed for harvesting small plots. After harvest, the grains of each wheat plot were individually weighed. Therefore, the unit of yield was gram per plot area (2.4 m$^2$). Since the plot size was identical for all plots in the three fields, yield is presented in terms of gram hereinafter.

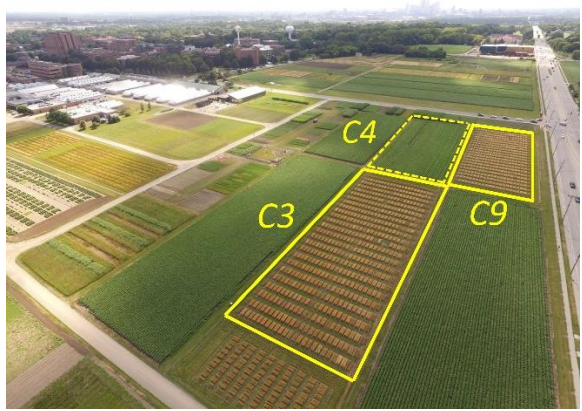

*Figure 1. Layout of the tree experimental yield trials during two consecutive growing seasons. Aerial imagery was performed from two cites (C3 and C9) in 2017 and one site (C4) in 2018 after crop rotation.*

## 2.2 Platform for aerial imagery

The UAV used in this study was DJI Matrice 600 Pro equipped with A3 Pro flight controller (Figure 2). Flight missions were created and executed in a grid mode with DJI Ground Station Pro. Table 1 presents the detail of the flight mission. For image collection, the entire mission was executed in autonomous mode except the take-off and landing which were performed manually. On the same day of image collection, a manual flight was performed at very low altitude (~ 5 meter) to collect images for endmembers extraction which is described in section 2.5.1.

A gimbal (DJI Ronin-MX) was used to carry the airborne hyperspectral imaging components and automatically maintain the camera at nadir position regardless of the UAV movements (Figure 2B).

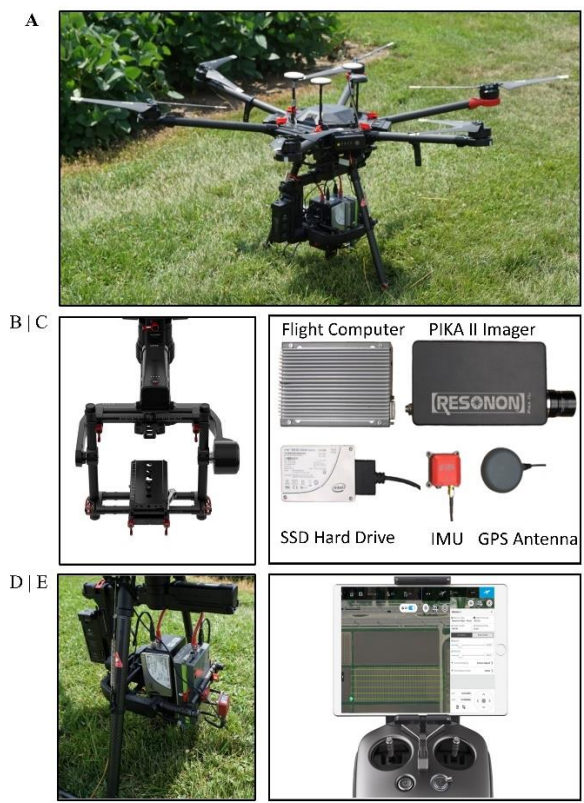

*Figure 2. (A) Unmanned aerial vehicle: DJI Matrice 600 Pro equipped with A3 Pro flight controller. (B) Gimbal: DJI Ronin-MX. (C) Components of airborne hyperspectral imaging system. (D) Airborne hyperspectral imaging system mounted on the gimbal. (E) Remote controller and DJI Ground Station Pro for creating flight missions.*

*Table 1. Flight information for two missions.*

| Flight mission for: | Flight mode | Altitude (m) | Speed (m/s) | Sidelap | Spatial resolution (cm) |
|---|---|---|---|---|---|
| Yield prediction | Autonomous | 20 | 2 | 50% | < 2 |
| Endmember extraction | Manual | ~ 5 | 0.5 | - | < 0.5 |

## 2.3 Airborne hyperspectral imaging setup

The camera used in this study was a push-broom hyperspectral camera (PIKA II, Resonon, Inc., Bozeman, MT 59715, USA) with the specifications presented in Table 2. The components of the airborne hyperspectral imaging system include the imager, flight computer, GPS antenna, inertial measurement unit (IMU), and a solid-state hard disk (Figure 2E).

The imager started collecting data as soon as the GPS of imaging system entered the predefined target area and the aircraft reached a defined minimum altitude (8 meters for yield prediction mission and zero for endmember extraction mission). To define the area of interest, a polygon was created with an appropriate buffer around the experimental field in Google Earth, saved as KML file, and then uploaded into the flight computer. The camera started image collection when the GPS of imaging setup entered the polygon above the predefined threshold, and it stopped the imagery once the aircraft exited from the polygon.

Image acquisition was performed in auto expose mode in which gain and exposure time were automatically adjusted based on the ambient lighting conditions and the brightness of the target.

The other user-defined parameter was the frame rate of scanning denoting the number of pixel lines scanned perpendicular to the direction of movement at each second. Frame rate of scanning was a function of two specifications of the imager including the field of view (i.e., 33 degree) and the number of spatial channels (i.e., 640) as well as two user-defined parameters including flight altitude and aircraft speed. In this study, a low flight altitude, 20-meter above ground level (AGL) was defined to attain a high spatial resolution while avoiding the potential turbulence over canopy caused by the propellers of UAV. The aircraft speed was set 2 m/s to cover the entire field in one flight. Once the flight altitude and speed were set, a frame rate of 108 frame per second was calculated as described by (Moghimi et al., 2017) to maintain the spatial integrity (square pixels with aspect ratio of 1:1 in cross and across track).

The hyperspectral pixel lines captured by PIKA II were transferred to the flight computer via an Ethernet cable, synchronized by GPS and IMU data, and then saved as a hyperspectral image cube to the hard drive through a USB-3 connection. With 2000 hyperspectral pixel lines collected per each image, the size of each hyperspectral image cube was 2000×640×240, requiring about 640 megabytes space for saving.

*Table 2. Specifications of PIKA II hyperspectral camera.*

| Hyperspectral imager | Spectral range (nm) | Spectral resolution (nm) | Spectral channels | Spatial channels | Maximum frame rate (frame per second) | Bit depth | Field of view (degree) |
|---|---|---|---|---|---|---|---|
| PIKA II | 400 – 900 | 2.1 | 240 | 640 | 145 | 12 | 33 |

## 2.4 Pre-processing of hyperspectral images

### 2.4.1 Radiometric Calibration

The hyperspectral images were collected as raw digital numbers (DNs) which is the least useful format with no units or physical meaning. Therefore, raw images were converted to radiance ($Wm^{-2}sr^{-1}nm^{-1}$) using the lab-derived radiometric calibration file provided by the manufacturer of imager. This conversion is a key step required for the radiometric calibration of hyperspectral images to compensate for the non-uniform spectral and spatial responses of the instrument (Moghimi et al., 2018b).

To account for potential variation in solar illumination, hyperspectral images in radiance were then converted to reflectance using reference panels (60×60 cm) placed in the field before image collection. The panels were pained with gray paint mixed with Barium Sulfate to diffuse the incoming solar irradiance in various directions (i.e., no specular reflection). In a laboratory setup, the actual reflectance of gray panels were measured by a ASD FieldSpec 4 spectroradiometer (Analytical Spectral Devices, Inc., Longmont, CO, USA) with respect to the reflection of a Spectralon panel (Labsphere, Inc., North Sutton, NH, USA) as a standard reference panel with highly Lambertian surface. Radiance and reflectance conversion were performed using SpectrononPro software (Resonon, Inc., Bozeman, MT 59715, USA). The gray panels were placed in alleys based on sensor footprint to maximize the probability of capturing at least one set of reference panel in each image. The unique ID of the plots located at both sides of gray panels were recorded in an inventory for further processing to recognize the ID of all plots across the image.

### 2.4.2 Noisy Band Removal

Prior to any further analysis, the first and last few bands were disregarded because of high noise (any bands before 430 nm and after 870 nm). In addition, spectral bands near the absorption region of $O_2$ and $H_2O$ were removed from the hyperspectral data cube (Moghimi et al., 2018b). In total, 190 spectral bands out of 240 bands were kept for further analyses.

### 2.4.3 Plot segmentation and identification

#### 2.4.3.1 Segmentation of plots from background

When aerial images were collected, wheat plots were at the senescence stage. While chlorophyll a and chlorophyll *b* in a green, healthy leaf of a wheat plant absorb a high extent of light at blue and red regions of electromagnetic spectrum for photosynthesis, a senescent leaf tends to absorb less light at these two regions – this is because of a significant decline in chlorophyll content (Lu et al., 2001). However, the extent of enhancement in reflection from senescent leaves of wheat at red region is higher than the reflection at blue region. The reason for this change in reflectance pattern is that carotenoid, with a high absorption at blue region (Lichtenthaler, 1987), is much less affected compared to chlorophyll a and b during leaf senescence, meaning the illuminated light is still highly absorbed at the blue region during senescence (Biswal, 1995; Grover et al., 1986). Based on this knowledge, a vegetation index referred to as *normalized difference plant senescence index* (NDPSI) was proposed in this study to segment wheat plots from background. NDPSI is essentially a vegetation index derived from two broad bands: red (670±5 nm) and blue (450±5), as follows:

$$NDPSI = \frac{Red - Blue}{Red + Blue} = \frac{\frac{1}{n}\sum_{i=665}^{675}\rho_i - \frac{1}{m}\sum_{j=445}^{455}\rho_j}{\frac{1}{n}\sum_{i=665}^{675}\rho_i + \frac{1}{m}\sum_{j=445}^{455}\rho_j} \quad (1)$$

where $\rho$ denotes reflectance at particular wavelength, $n$ and $m$ refer to the number of bands used to generate broad red and blue spectral bands ($n = m = 5$), respectively. While single bands at 670 and 450 nm can be also used to calculate NDPSI, consolidating five bands as broader red and blue bands rendered a NDSPI gray-scale image with effectively reduced salt-and-pepper noise.

Pixels representing wheat plots displayed a tendency to exhibit large values of NDPSI compared to the background pixels, which were mainly reference panels, green winter wheat planted in alleys, soil, and shadow caused by plants. A threshold was defined for pixel values of NDPSI to segment wheat plots from the background (Figure 3). Afterwards, several morphological operations were applied on this binary image. First, a flood-fill operation was conducted to fill the holes generated inside the objects. The second mathematical morphology was opening operation, an erosion followed by a dilation with a rectangle structuring element (10×5), to remove small objects and break the potential connection between adjacent plots due to the lodging of plants. To assure small objects are disregarded, a threshold was defined for the area of the objects in terms of pixels. The obtained binary mask was then used to segment the plots and fit bounding boxes enclosing the plots (Figure 3).

#### 2.4.3.2 Recognizing plots ID

The geo-rectification process of hyperspectral images failed largely because the IMU data was not accurate enough due to the magnetic interference. Therefore, a semi-automatic pipeline was developed to identify the plot ID of segmented plots in each image.

*Manual processing*

For each hyperspectral image, the ID of two plots next to the gray panel were identified through the following steps. GPS/IMU data was utilized to generate an approximate swath outline (i.e., image boundary) per each image. These swath outlines were saved as KML files and imported into QGIS (QGIS Geographic Information System, 2018) to have an estimate for the geographical position of the field area scanned in each hyperspectral image. To create a detailed basemap with reference features, such as the location of reference panels, an orthomosaic was generated using high-resolution RGB images orthorectified and stitched together using Pix4Dmapper (Pix4D S.A., Lausanne, Switzerland). The RGB images were captured at low altitude (7 meter) by DJI Inspire UAV equipped with a double 4K sensor (Sentera, Inc., Minneapolis, MN 55423, USA). Using the orthomosaic as the basemap for the swath outlines assisted in identification of at least two plots per image since the plot ID of the plots located at both side of gray panels was previously recorded.

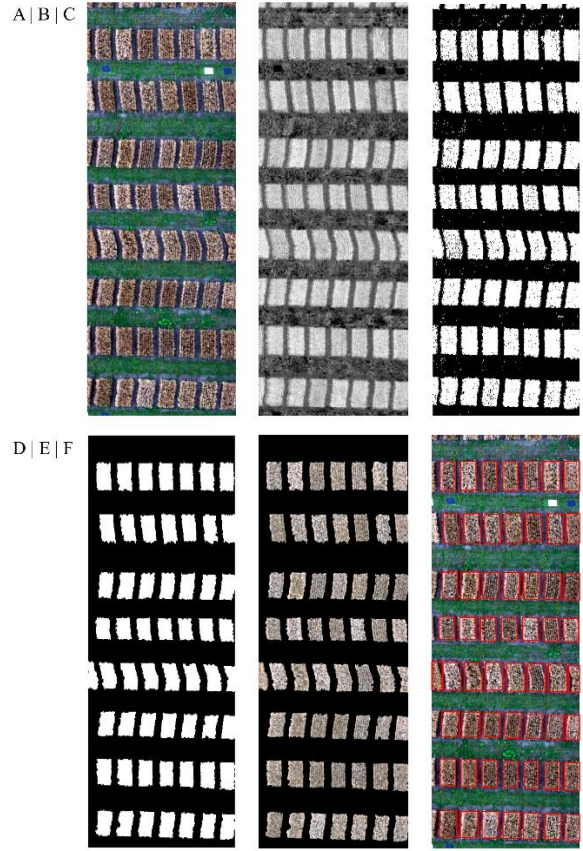

Figure 3. (A) RGB representation of a hyperspectral image. (B) Gray scale image of normalized difference plant senescence index (NDPSI). (C) Binary image obtained by thresholding NDPSI. (D) Binary mask obtained by morphological operations including flood-fill, opening, and area thresholding. (E) RGB representation of hyperspectral image of plots segmented from background using the binary mask. (F) Fitting bounding boxes enclosing the segmented wheat plots.

*Automatic processing*

An algorithm based on image processing techniques was developed to assign a plot ID to the plots in each hyperspectral image using one of the two plots next to the gray panel as a reference. Pixels representing the top-left corner of bounding boxes in a single hyperspectral image were clustered based on row- and column-wise pixel distance from each other (Figure 4). The number of clusters obtained by row- and column-wise clustering denoted the number rows and columns of plots exist in a given image, respectively. Afterwards, a grid was created from horizontal and vertical lines obtained by taking the average over the row as well as column arrays of the pixels (i.e., top-left corner of bounding boxes) grouped in one cluster (Figure 4C).

The process of assigning a plot ID to each grid cell was automatically propagated across the image using one of the two reference plots next to the gray panel along with the plots ID map, which included the ID and location of the plots relative to each other in the field. Theoretically, each cell of the grid should entail a wheat plot with a unique ID, and potentially a bounding box encompassing the plot. However, a few plots were not detected by the segmentation algorithm since they could not pass the thresholds defined in morphological operations. If a bounding box existed in a given cell gird, the algorithm allocated the ID assigned to the cell to the bounding box. If there was no bounding box in a given cell, then the algorithm skipped to the next grid cell.

Subsequent to plot segmentation and identification, plots were cropped from the hyperspectral images using the fitted bounding boxes and saved as 3-D matrices ($x \times y \times \lambda$) to preserve the spatial ($x \times y$) and spectral ($\lambda$) integrity of plots for further workflow. These 3-D matrices will be referred to as plots hyperspectral cube (P-HSC) hereinafter.

## 2.5 Hyperspectral image analysis

### 2.5.1 Endmember selection

Despite the high-spatial resolution (~2 cm) attained by flying at 20-meter altitude, each pixel might exhibit spectral characteristics of a mixed pixel, largely due to properties of the objects of interest (spikes and leaves) such as size, angle, and curvature. For instance, with the spatial resolution of 2 cm, it was rather infeasible to find a pixel that contains only a spike because of the spike geometry from the sensor perspective. To obtain a sufficient resolution for capturing pure spectral signatures, called endmember, representing the objects in the hyperspectral image dataset, a low altitude flight (5-meter AGL) was performed in a manual mode – the attained spatial resolution was approximately 0.5 cm.

It should be noted that the notion of endmember existence in the form of perfectly pure pixel is for conceptual convenience because of uncertainty caused by sensor noise and spectral signature variability within a class (Schowengerdt, 2012). In practice, each pixel is essentially a mixed pixel to a certain extent in remote sensing. Therefore, the most pure pixels in the scene with the most distinct spectral response were considered as the endmembers.

In hyperspectral image datasets, there were six distinct classes, including spikes, wheat leaves, soil, shadow, winter wheat, and gray panel. Therefore, the spectral response of a pixel can be composed of these six classes, each contributing with various extent and with a distinct spectral signature. To distinguish the abundance of these classes in each pixel of the images collected at 20-meter altitude, six endmembers, each representing a single class, were identified form the images collected at 5-meter altitude.

One of the widely-used techniques to identify the endmembers is N-FINDR algorithm, in which $n$ endmembers are selected as the $n$ vertices of a $(n-1)$-simplex with a maximum volume encompassing the majority of pixels in the feature space spanned by all pixels (Winter, 2004, 1999). However, the N-FINDR algorithm suffers from issues such as long processing time, and inconsistency in selecting the final set of endmembers due to the random initial endmember selection. Various automated techniques, all inspired by N-FINDR, were proposed to ameliorate the process of endmember extraction (Chan et al., 2011; Chang et al., 2011; Zortea and Plaza, 2009). In the present study, successive volume maximization (SVMAX), proposed by Chan et al. (2011), was utilized to identify the endmembers through a successive optimization problem. The number of endmembers in SVMAX was set to six, as there were six distinct classes.

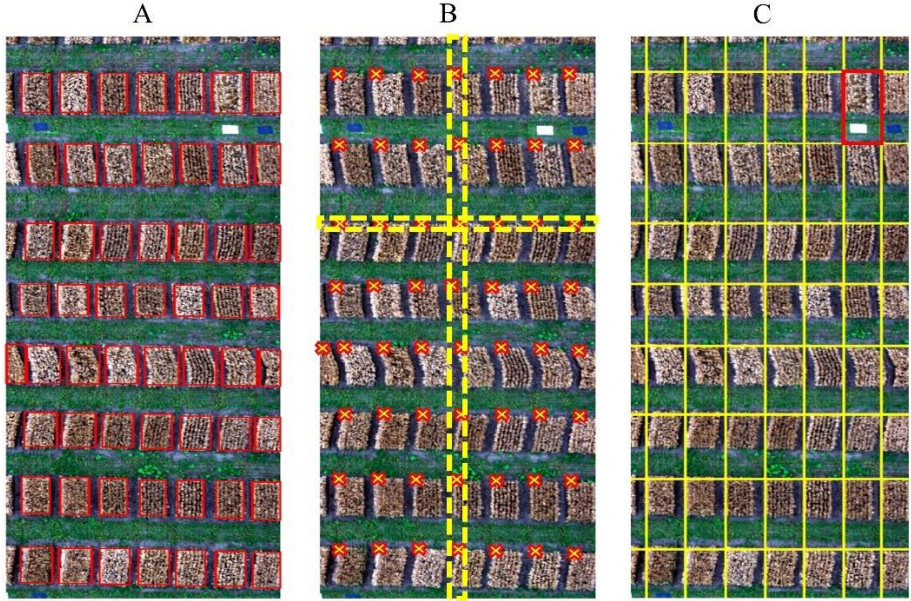

*Figure 4. (A) Bounding boxes enclosing the wheat plots. (B) Top-left corner of bounding boxes are clustered based on row- and column-wise pixel distance. It is shown how top-left corners are clustered together in a row- and a column-wise manner. In this example, there are eight rows and eights columns of plots (C) The row array of pixels in a row cluster, and the column array of the pixels in a column cluster were averaged to generate the horizontal and vertical lines, respectively. Based on the plot ID of the plot next to the gray panel (red box) along with the plots map, a plot ID was assigned to the grid cells, and subsequently to the bounding boxes inside them.*

### 2.5.2 Spectral Mixture Analysis

Once the endmembers were identified from the images captured at low altitude, each pixel of a P-HSC can be represented as a convex combination of the endmembers. Since P-HSCs mainly contained spikes, wheat leaves, soil, and shadow, only four endmembers representing these four classes were used for the un-mixing process. In this study, to determine the fractional abundance of the endmembers in the pixels of P-HSCs, a matrix factorization problem with two constraints was defined as per Thurau et al. (2010) in which a Frobenius norm is minimized as follows:

$$min \, \|X - WH\|_F$$
$$s.t. \begin{cases} 1^T \cdot h_j = 1 \\ 0 \leq h_{ij} \leq 1 \end{cases} \quad (2)$$

where $X(d \times N)$ is the matrix of data obtained by reshaping a P-HSC (i.e., 3D matrix) to a 2-dimenionsal matrix such that pixels ($N$: number of pixels) were extracted in column-wise order and were placed as the columns of matrix $X$, and bands ($d$: number of bands) were placed as the rows. $W(d \times e)$ and $H(e \times N)$ are the endmembers matrix ($e$: number of endmembers), and the abundance matrix, respectively. Each column ($h_j$) of matrix $H$ was calculated by resolving a quadratic optimization problem (Moghimi et al., 2018b) iteratively $N$ times with constraints similar to equation 2 as follows:

$$\min_{h_j} \, \frac{1}{2} h_j^T Q h_j + c^T h_j, \quad j = 1, \ldots, N$$
$$s.t. \begin{cases} 1^T \cdot h_j = 1 \\ 0 \leq h_{ij} \leq 1 \end{cases} \quad (3)$$

where

$$Q = 2W^T W$$
$$c = -2W^T x_j \quad (4)$$

### 2.5.3 Sub-plot image analysis

The distribution of the measured yield for a plot was not homogeneous over the plot because of the factors such as spatial variability of soil, available nutrient, and marginal effects. While studying the yield variation within a plot can provide valuable insights into the breeding program for selecting advanced wheat lines, harvesting the wheat grains at sub-plot resolution in a large yield trial is a tedious, unrealistic, and impractical task. In this study the high spectral and spatial resolutions of aerial hyperspectral images were leveraged to examine the yield variation within a plot.

Each plot was divided into square sub-plots (15×15 pixel). To assure that P-HSC can be divided into 15×15 grids, zero-padding was applied at the margins of P-HSC, meaning each pixel can be fitted in a 15×15 grid. Once a plot was divided into sub-plots, a yield should be assigned to each sub-plot. For this purpose, we hypothesized that the yield of a sub-plot is proportion to the number of spikes and leaves (SL) pixels which are representing the above-ground biomass in that subplot (i.e., a subplot with higher density of spikes and leaves contributes more in the plot yield). To count the number of SL pixels within each sub-plot, sub-plot pixels were classified into two classes: SL class or soil-shadow (SS) class. A given pixel was classified to SL class if the summation of abundance for spikes and leaves endmembers in that pixel was more than 0.5; otherwise, it was assigned to SS class (background). Afterwards, given the measured yield of a plot was $y$, a normalized yield value $y_i$ was assigned to the $i$th sub-plot as follows:

$$y_i = \frac{n_i}{N} \times y \qquad j = 1, \dots, m \qquad (5)$$

$$N = \sum_{i=1}^{m} n_i \qquad (6)$$

where $n_i$ is the number of SL pixels in $i$th sub-plot within a given plot, $N$ is the total number of SL pixels in the plot, and $m$ is the number of sub-plots in the plot. Since the yield assigned to a sub-plot was normalized based on the total number of SL pixels in the plot, the summation of yield for all sub-plots within the plot was equal to the measured yield for the plot ($\sum_{i=1}^{m} y_i = y$).

### 2.5.4 Extracting input features from sub-plots

Each sub-plot was composed of several SL pixels segmented from SS pixels. These SL pixels were considered as one object per each sub-plot window. Object-based image analysis (OBIA) approach was then used to leverage extracting features (such as size, area, texture, mean and standard deviation per band) associated with a set of pixels as opposed to per-pixel analysis (Blaschke, 2010).

In the present study, mean and standard deviation (std) per band (in total 190 bands) were extracted as the input features because they offered adequate information to estimate the distribution of pixels' reflectance per band per each subplot. The other input feature extracted from the sub-plots was the area of the SL object in terms of pixels (i.e., the number of SL pixels). This refers to the number of samples used to calculate the mean and std of the distribution. In total, the number of input features per sub-plot was 381 (190+190+1).

## 2.6 Dataset

There were three sets of data from adjacent fields C3 and C9 collected in 2017, and C4 collected in 2018. After removing the damaged plots, a set of 50 plots was selected as the test dataset using stratified sampling to assure that the test dataset has an akin yield distribution to the training and validation datasets (Figure 5; Table 3). The sub-plots of these 50 plots were held out as the test dataset for an unbiased evaluation of the final trained model. Subsequent to the test dataset selection, other plots of the three fields were divided into sub-plots and merged together to form a dataset for training and validation of the model. Using stratified sampling, these sub-plots were split into training (90%) and validation (10%) datasets to train and validate the model during the training process (Table 3).

In another experiment, an individual model was developed per each field. With a similar approach described above, the dataset of each field was separately divided into training (85%), validation (15%) after keeping aside the sub-plots of 50 plots selected for test datasets.

After splitting the data, the training dataset was normalized to make each feature have zero-mean and unit-variance. Subsequently, validation and test datasets were standardized using the mean and variance obtained from training dataset.

*Table 3. Number of plots and sub-plots in each field and size of training, validation, and test datasets for three individual models developed for each field as well as the model trained on the large training dataset obtained by merging all three fields.*

| Year | Field | Number of plots | Number of sub-plots | Number of plots for test dataset | Number of sub-plots for test dataset | Number of sub-plots for training dataset | Number of sub-plots for validation dataset |
|---|---|---|---|---|---|---|---|
| 2017 | C3 | 422 | 19287 | 50 | 2239 | 14491 | 2557 |
|  | C9 | 345 | 19650 | 50 | 2776 | 14343 | 2531 |
| 2018 | C4 | 254 | 12773 | 50 | 2507 | 8726 | 1540 |
| All fields |  | 1021 | 51710 | 50 | 2530 | 44261 | 4919 |

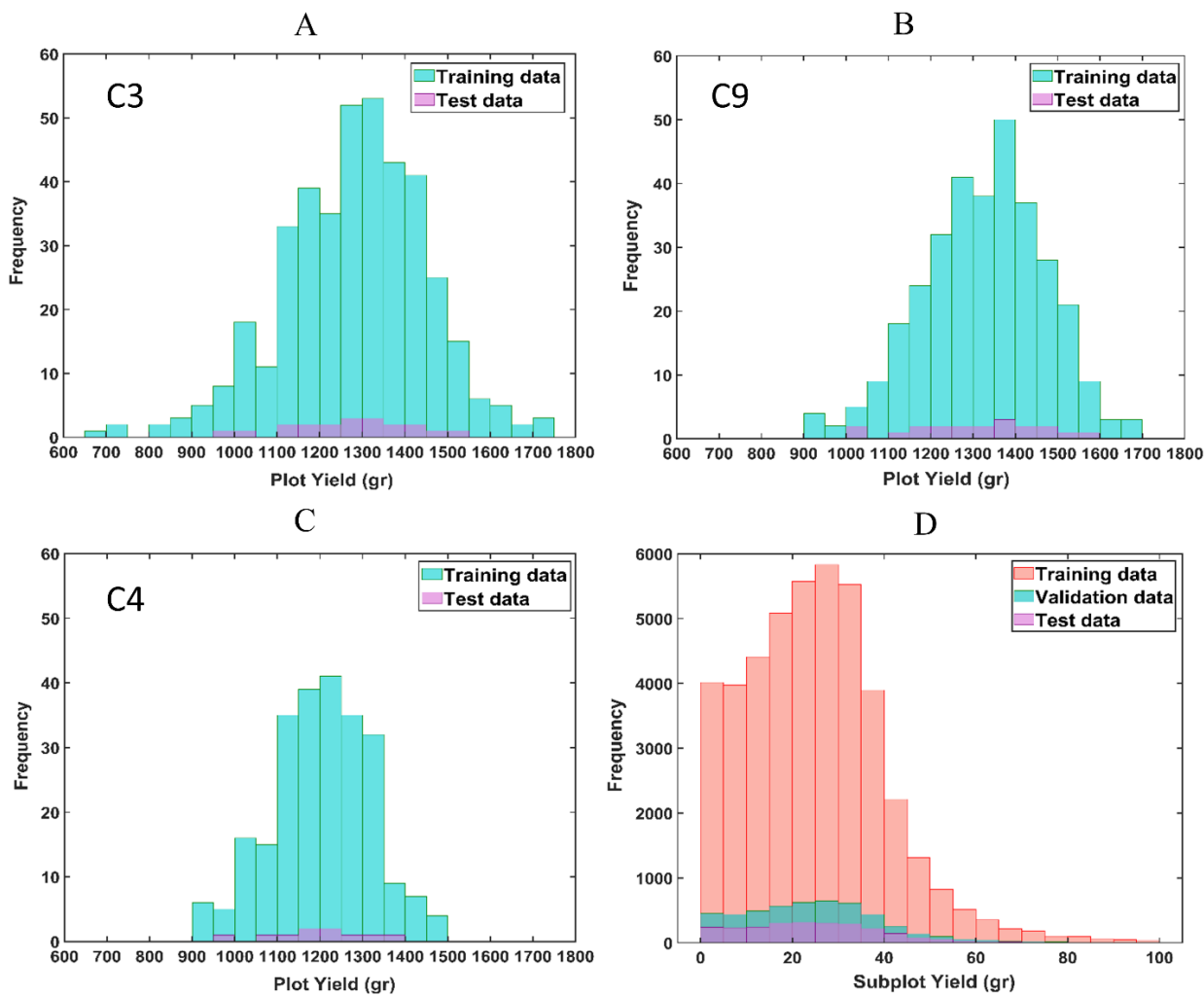

*Figure 5. Yield histogram of the plots used for training and test per each field for developing a model based on merged datasets. For the test dataset, 20 plots from C3 (A), 20 plots from C9 (B), and 10 plots from C4 (C) were selected. The number of plots selected for the test dataset from each field were proportion to the total number of sub-plots belonging to that field. (D) Yield histogram of the sub-plots used for training, validation, and test datasets.*

## 2.7 Deep Neural Network

Among various type of deep learning architectures, convolutional neural network (CNN) (Krizhevsky et al., 2012; LeCun et al., 1990, 1989) is well suited for data with spatial structure such as image-based datasets. However, the spatial information within sub-plots was lost because the yield assigned to the sub-plots was based on the number of SL pixels, regardless of the spatial location of SL pixels with respect to each other in the sub-plot window. Consequently, a vector of features for each sub-plot was considered as the input layer for a deep neural network (DNN) with fully connected layers in preference to CNN. In this study, the network was a feedforward neural network, also known as multilayer perceptron (MLP) (Goodfellow et al., 2016), composed of an input layer, an output layer, and four hidden layers.

The input layer represented the input features. Since 381 features were extracted from the sub-plots, the input layer had 381 units (Figure 6). The output layer was a single unit representing the predicted yield. The number of hidden layers and their units were two important hyper-parameters of the network defined through an empirical process in which the performance of various network architectures, selected based on the domain knowledge, were evaluated. Since a large portion of wavelengths scanned by the hyperspectral camera are redundant or irrelevant to the desired phenotyping trait (Moghimi et al., 2018a), the number of units in the hidden layer was selected among a set of small numbers compared to the input layer. Alternatively, the number of hidden layers was limited by the size of the training dataset because an additional hidden layer increased the required number of data to train the model parameters (weights and biases).

Layers were fully connected, meaning each unit in layer $l$ was connected via weighted linkage to all units in the layer $l + 1$. Therefore, the input of unit $j$ in layer $l + 1$ (denoted by $z_j^{l+1}$) was the weighted sum of the output of all units in layer $l$ ($a_i^l$) plus a bias term ($b_j$), and the output of the unit $j$ in layer $l + 1$, denoted by $a_j^{l+1}$, were calculated as follows:

$$z_j^{l+1} = \sum_{i=1}^{d} w_{ij}^l \, a_i^l + b_j^l \quad (7)$$

$$a_j^{l+1} = f(z_j^{l+1}) \quad (8)$$

where d is the number of units in layer $l$, $w_{ij}^l$ refers to the weights of links connecting all the units in layer $l$ into unit $j$ in layer $l + 1$, and $f(.)$ was the activation function to introduce non-linearity into the network. The transfer function was rectified linear unit (ReLU) $f(x) = \max(0, x)$ (Glorot et al., 2011; Jarrett et al., 2009; Nair and Hinton, 2010), a non-linear function that allows the network to learn faster and avoids saturation for large positive inputs.

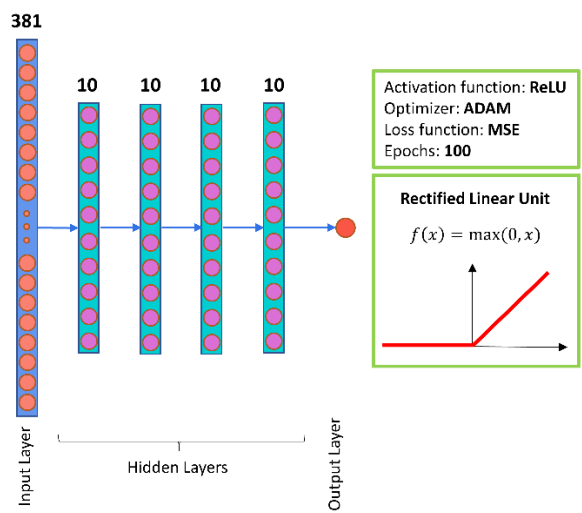

*Figure 6. The architecture of deep neural network with fully connected layers.*

The weights and biases were adjusted through an iterative process using training and evaluation datasets. The training began with initial random values as per Glorot & Bengio, (2010) for weights and biases. In the forward phase of an iteration, weights and biases were used to calculate the network output using equations (7) and (8). In the backward phase, the weights and biases were updated based on the network error. The cost function used to measure the network error was mean squared error (MSE) which is the squared difference between the output of the network (i.e., the predicted yield for a sub-plot) and the desired value (i.e., the yield value assigned to the sub-plot), averaged across all training samples.

During the training process, an optimization algorithm was used to identify the network parameters for which the cost function was minimized (returned a low value for the error). The optimization algorithm was Adam (Kingma and Ba, 2014), a computationally efficient optimization algorithm with an adaptive learning rate. Adam embraces the benefits of adaptive gradient algorithm (Duchi et al., 2011) and root mean square propagation (Tieleman and Hinton, 2012), which are suitable for sparse gradients and non-stationary setting, respectively (Kingma and Ba, 2014). The number of times that the entire dataset was passed through the network, known as epoch, to find optimized values for parameters of the network was set to 100.

### 2.8 Computational environment

The DNN model was developed and tested in Keras 2.2.2 (Chollet and others, 2015) with TensorFlow 1.9.0 (Abadi et al., 2015) backend running on an NVIDIA (GeForce GTX 750 Ti) GPU. All other computations and image analysis were performed by MATLAB R2017b (MathWorks, Inc., Natick, MA, USA).

## 3 Results

### 3.1 Endmember extraction

Pixels identified by SVMAX as the endmembers were the vertices of a simplex with the maximum volume compared to any other possible simplex formed by pixels in the feature space spanned by all pixels. To account for uncertainty caused by factors such as sensor noise, the reflectance of pixels within a specified Euclidean distance of the identified endmembers were averaged as the new set of endmembers. For visualization of the endmembers location with respect to the other pixels, all pixels were projected onto a 2- and 3-dimensional feature space, respectively spanned by the first two and three principal components (PC) obtained by principal components analysis (Figure 7). It should be noted that the location of endmembers might not be the vertices in the new feature space because of projection onto a lower dimension. For instance, in a 2-dimensional feature space, the endmembers of gray panel, winter wheat, and shadow were the vertices of a triangle while the endmembers of spike, leaves, and soil were placed inside the established triangle. Alternatively, in a 3-dimensional feature space spanned by the first three PCs, a different set of endmembers might be the vertices depending on the viewing angle (Figure 7).

Figure 8 illustrates the spectral signature of the endmembers. Based on the spectral signature of endmembers and configuration of endmembers' location in the 3-dimensional feature space, it can be inferred that spectral response of spikes and senescence leaves as well as soil and shadow tend to be similar, whereas, winter wheat had the most distinct spectral signature among all six endmembers.

### 3.2 Spectral un-mixing

The spectral response of four endmembers, including spikes, leaves, soil, and shadow, were used for un-mixing analysis of P-HSC because the pixels representing winter wheat and gray panels were masked out during the segmentation process (Figure 9). The quadratic optimization problem, defined to minimize the Frobenius norm, returned four gray scale images, each of which representing the abundance of a particular endmember (Figure 9B). Therefore, for a given pixel in a P-HSC, there were four values denoting the abundance of endmembers such that the summation of these four values was equal to one due to the applied constraints in solving the optimization problem. To segment pixels representing biomass (i.e., SL class), the abundance of spikes and leaves were added pixel-wise. A binary mask was created to segment SL pixels (Figure 9C). A pixel was assigned to SL class if the summation of spikes and leaves abundances was more than the summation of soil and shadow abundances.

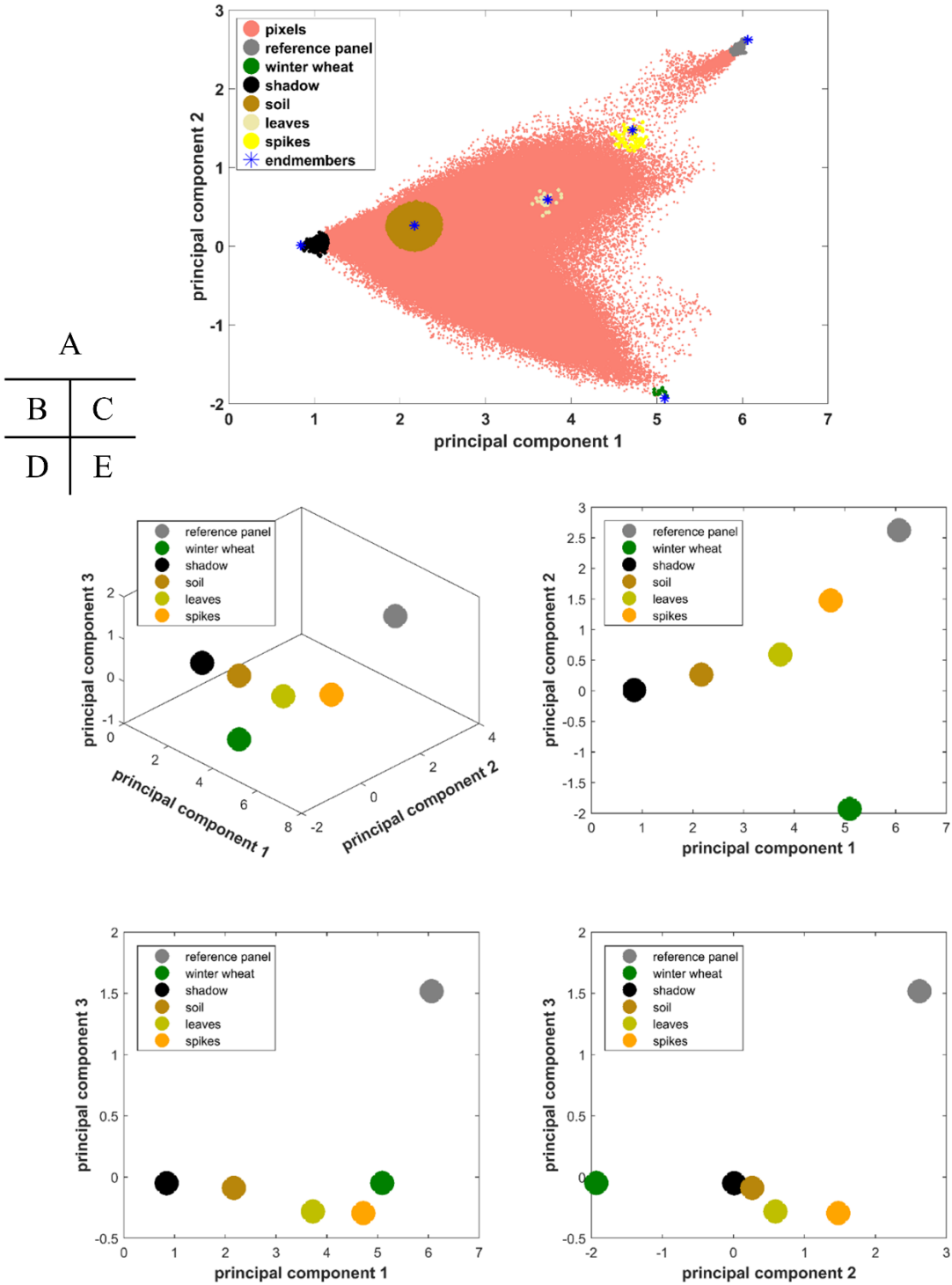

*Figure 7. Location of the endmembers in the feature space spanned by the first two principal components (PC) (A), and the first three PCs (B). (C) Projecting the endmembers on the PC1 and PC2 plane. (D) Projecting the endmembers on the PC1 and PC3 plane. (E) Projecting the endmembers on the PC2 and PC3 plane. Depending on the projection, different set of endmembers become the vertices.*

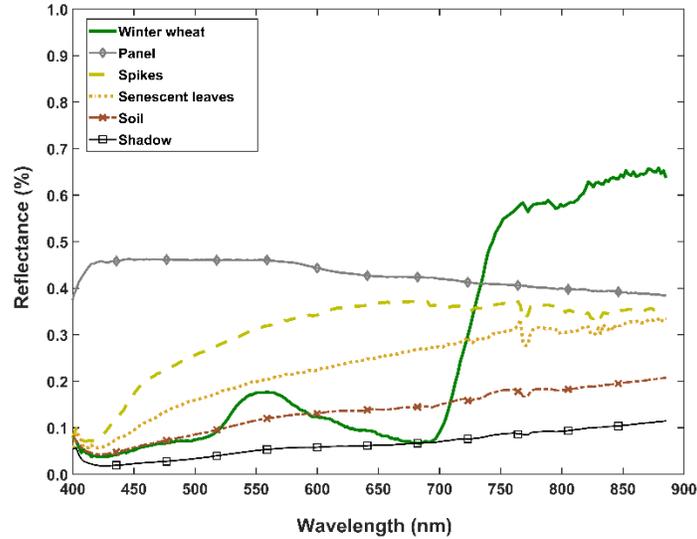

*Figure 8. Spectral response of the six endmembers.*

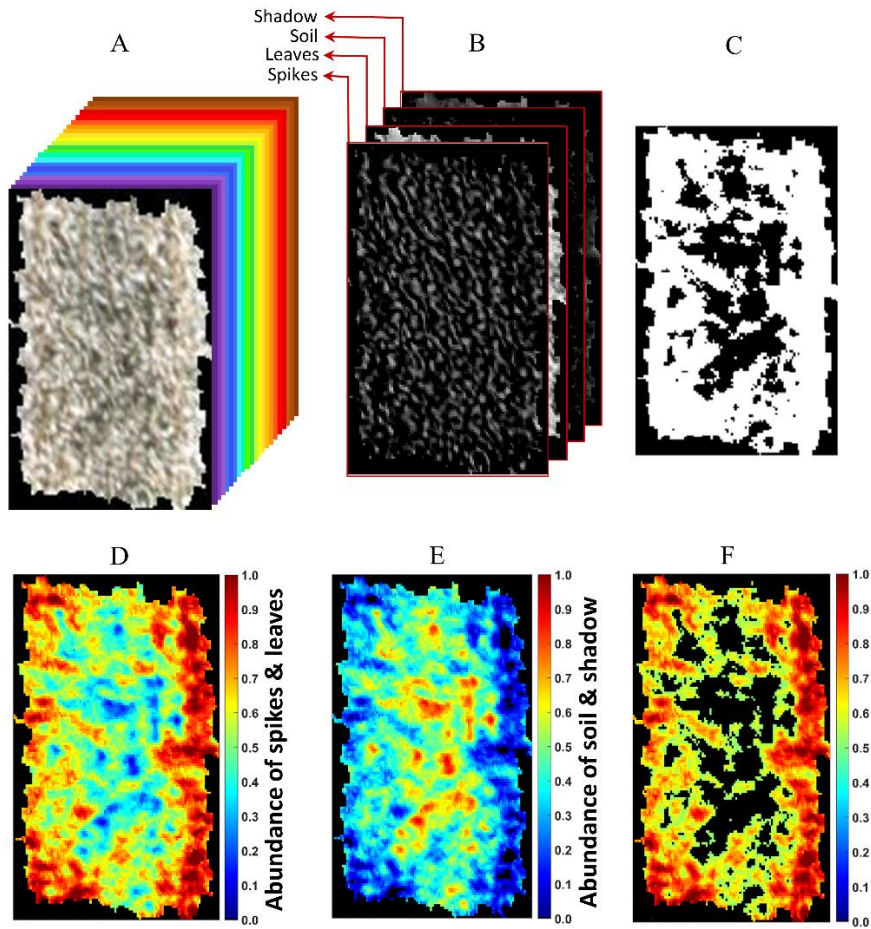

*Figure 9. (A) Hyperspectral cube of a plot (P-HSC). (B) Abundance of endmembers in each pixel shown as gray scale images. (C) Binary mask of spikes and leaves class. A threshold of 0.5 was applied on the summation of spikes and leaves abundances. (D) Summation of spikes and leaves abundances for each pixel shown as a colormap. (E) Summation of soil and shadow abundances for each pixel shown as a colormap. (F) Spikes and leaves pixels (SL class) were masked from the background.*

## 3.3 Yield allocation to sub-plots

The measured yield for a plot was distributed among the sub-plots based on the ratio between the number of SL pixels in sub-plots to the total number of SL pixels in the plot. Each sub-plot represents an area about 30×30 cm on the ground because the size of sub-plot was 15×15 pixels and the size of pixels were about 2 cm.

Several sub-plot window sizes were evaluated to find an appropriate window size. While a small window size allowed investigating the yield variation at a higher spatial resolution, the allocated yield to the sub-plots became very small as the number of SL pixels in sub-plot windows decreased. In addition, the probability of having sub-plots with the same number of SL pixels increased, meaning an identical yield was assigned to a significant portion of sub-plots within a given plot. This could deteriorate the process of training the model since a significant portion of sub-plots had identical target variables. For instance, by dividing a plot shown in Figure 10 with a window size of 10×10, more than 40 sub-plots had yield values varying between 12.5 and 15 grams (Figure 11), and more than 54 percent of sub-plots had identical yield. Alternatively, for a larger window size, the assigned yield to sub-plots varied substantially at the cost of sacrificing the spatial resolution for investigating the yield variation within a plot. By dividing the same plot shown in Figure 10 using a window size of 20×20, the yield of sub-plots varied from zero to about 60 grams with only about 8% identical sub-plots yield. To maintain the possibility of investigating yield variation at a higher spatial resolution and avoid numerous sub-plots with identical yield, the size of window was set to 15×15, compromising the benefits of 10×10 and 20×20 window size.

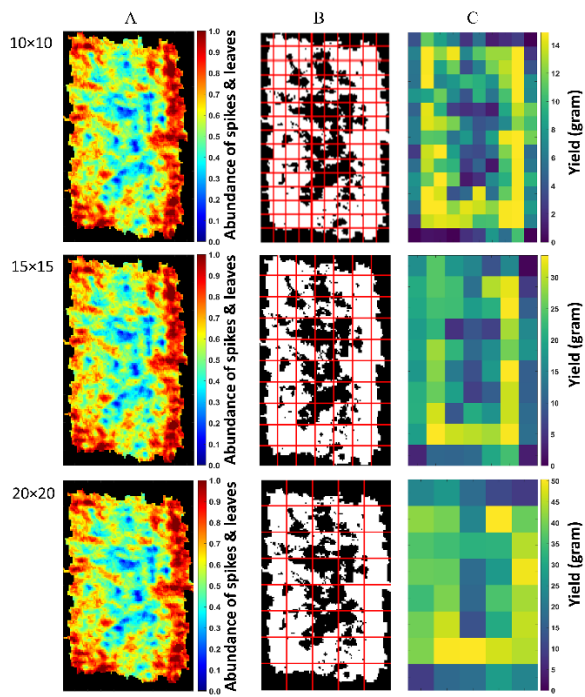

*Figure 10. Dividing a plot using various window size of 10×10, 15×15, and 20×20. (A) Summation of spikes and leaves abundances in a plot. (B) Dividing the binary mask of spikes and leaves into sub-plots using various window sizes. (C) Yield allocated to the sub-plots shown in colormap.*

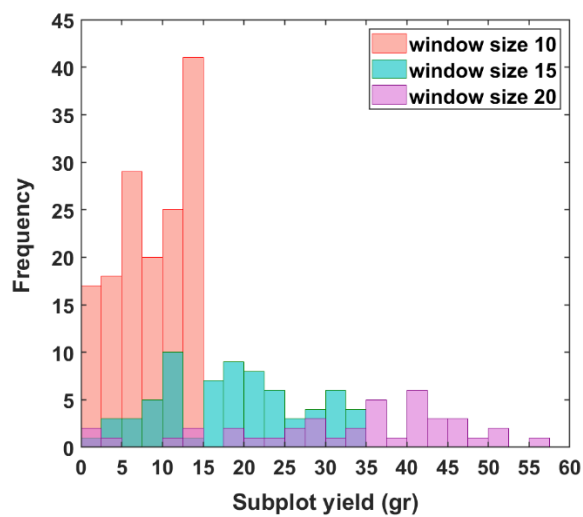

*Figure 11. Yield histogram of sub-plots generated by various window sizes. For the window size of 10×10, the variation range of the allocated yield values to the sub-plots was small, whereas the variation range of the allocated yield values to the sub-plots generated by a window size of 20×20 was wide at the cost of sacrificing the spatial resolution for investigating the yield variation within a plot.*

## 3.4 Deep Neural Network

### 3.4.1 Yield prediction at sub-plot scale

The training dataset of sub-plots was used to train a DNN model. In training the model, the main goal was to identify a set of model parameters (weights and biases) that minimize the cost function's value (i.e., RMSE). As training continued, model parameters were updated. To achieve an interpretable unit (gram) as the target value (yield), root mean square error (RMSE) was calculated for presenting the variation of cost function over epochs. Figure 12 illustrates how RMSE changed over training epochs for three individual models developed for each field as well as the model trained on the large training dataset obtained by merging all three fields. For all four models, RMSE decreased rapidly over the first training epochs for both training and validation datasets and, subsequently, reached a plateau where RMSE remained rather unchanged. However, for the merged dataset, there was a sharp decrease in RMSE within the first few epochs, meaning that the convergence occurred faster than other models. Among the 100 epochs, the weights and biases returning the lowest RMSE for validation dataset were saved as the model parameters to predict the yield of test dataset.

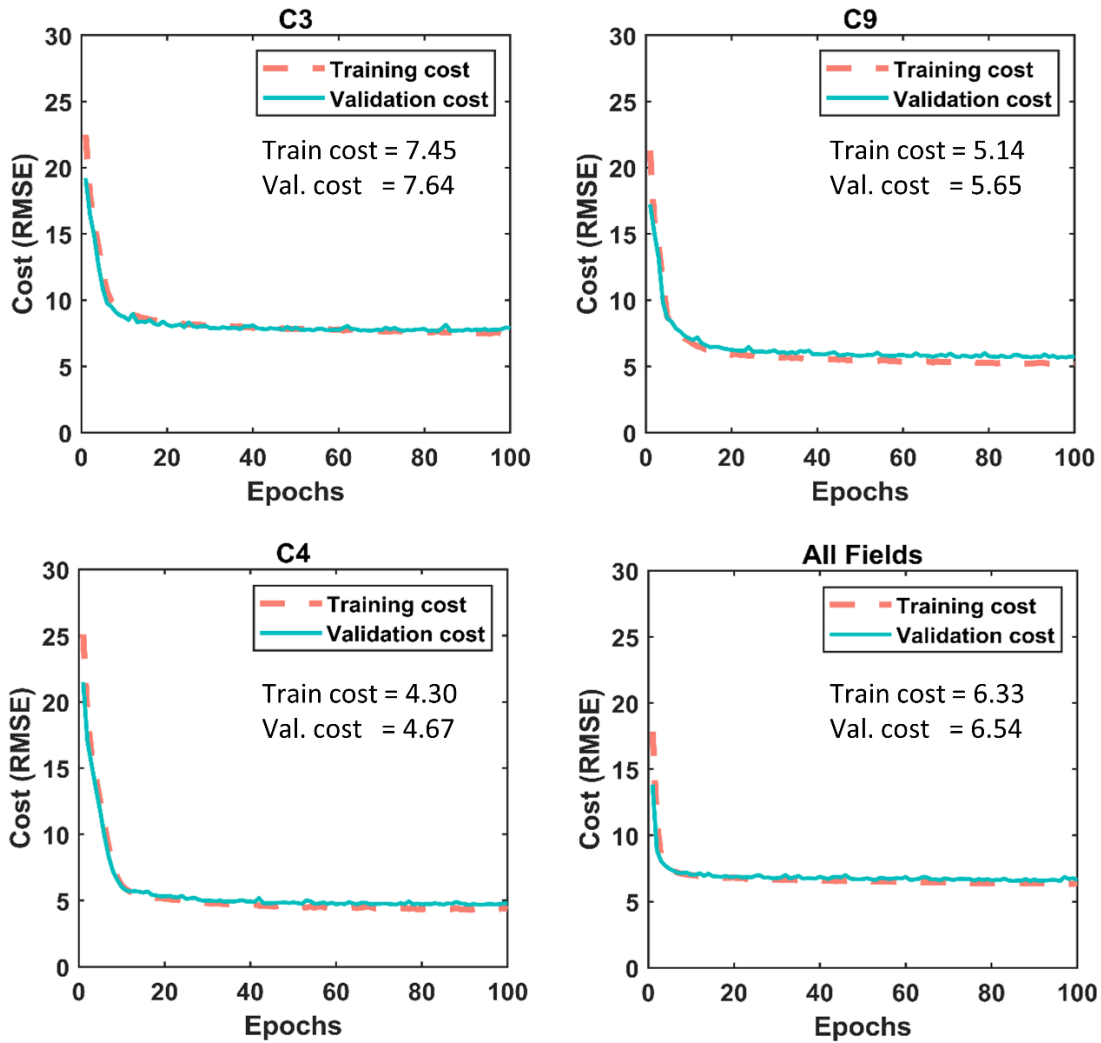

*Figure 12. Variation of root mean squared error over epochs for C3, C9, C4, and merged dataset.*

Figure 13 demonstrates the performance of the trained models in predicting the yield of the sub-plots in the test datasets. The model trained on the C9 dataset had the largest coefficient of determination ($R^2$) and lowest RMSE in predicting the yield. Alternatively, the C3 model had the lowest $R^2$ and largest RMSE, indicating the generalization of the trained model on an unseen dataset was not as satisfactory as the C9 model. This could be anticipated because the train and validation cost for the C3 model during the training process was the largest among the models (Figure 12). One reason that might explain the difference in performance of models ($R^2$ and RMSE) among the fields is the difference between the dates that images were captured from these two fields in 2017. The time interval between imagery and harvesting of C3 was one week more than C9. This suggests that the aerial imagery performed closer to the harvesting time might have a better correlation to the actual yield.

The model trained on the merged dataset had a promising $R^2$ ($R^2 = 0.79$) and low RMSE of 5.90 grams, indicating the DNN model could explain 79 percent of the yield variation among the 2530 subplots in the test dataset. The DNN model was able to predict the yield of a significant portion of the test sub-plots with a low error as they located nearby the 1:1 line (the black dashed line in Figure 13). However, the model demonstrated a tendency to underestimate the yield of sub-plots with the large yield value (more than 40 grams). This might be because the number of sub-plots having a yield of more than 40 grams in the training dataset was substantially lower than the number of sub-plots with the yield less than 40 grams (Figure 5D). Therefore, the network was moderately successful to learn the yield estimation based on the input data with a large sub-plot yield. A similar pattern was observed in yield prediction of the individual fields (Figure 13).

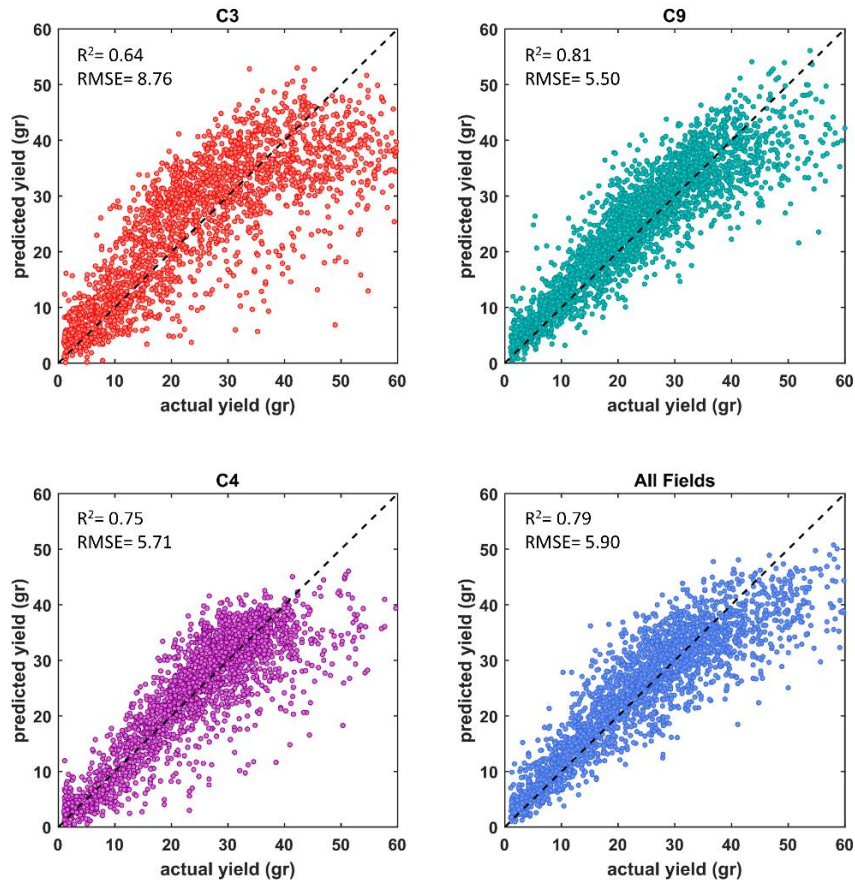

*Figure 13. Performance of deep neural network models on yield prediction of sub-plot test datasets. The model trained on the C9 training dataset had the best performance ($R^2=0.81$ and RMSE=5.5 grams), while the model trained on the C3 had the lowest $R^2$ and largest RMSE.*

### 3.4.2 Yield production in the middle one-third of the plot

An additional analysis was performed to determine the potential of each wheat variety in producing yield in the middle one-third of the plot. The results suggest that about 90 percent of the plots produced more yield at one side of the plot (Figure 14). Lower productivity in the middle one-third can be a result of the high competition between the plants in the middle of the plot and/or receiving less light compared to the plants at the border of the plots. Investigating the yield production capability of wheat lines in the middle one-third of the plots is of great interest to breeders as an extra valuable information. Breeders can use this as a decision-making tool to determine wheat lines incapable of producing sufficient yield in the middle one-third of the plot, where there is more competition, and eliminate them form their breeding program.

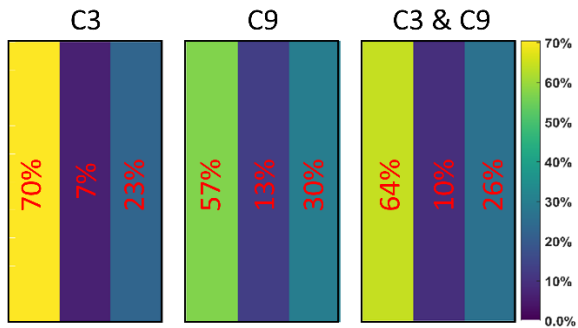

*Figure 14. Analysis of yield production in the middle one-third of the plots in C3 and C9 fields. About 90 (64+26) percent of the plots produced more yield at one side of the plot because of receiving appropriate extent of light and less competition for water and nutrient.*

### 3.4.3 Yield prediction at plot scale

As described before, the test sub-plots were obtained from the 50 test plots selected from the three fields using stratified sampling. To observe the performance of the trained model in predicting the yield at plot scale, the summation of the predicted yield for the sub-plots belonging to a test plot were compared to the measured yield for that plot (Figure 15). For yield prediction at plot scale, $R^2$ dropped to 0.41 compared to the $R^2$ of 0.79 for the sub-plots scale. To compare the prediction error obtained for plot and sub-plot analysis and account for the difference in the scale of yield variation, the normalized RMSE was calculated by dividing the RMSE of plot and sub-plots to their mean of yield. The normalized RMSE for predicting the yield at plot level was 0.14, while it was 0.24 for yield prediction of sub-plots, indicating the error in yield prediction of plots improved although $R^2$ deteriorated compared to the sub-plot scale.

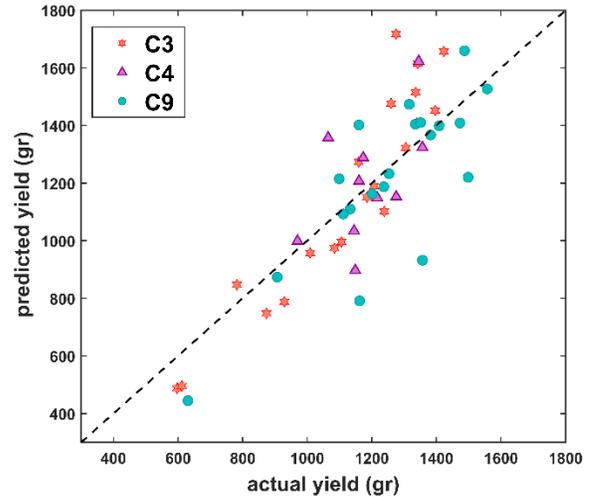

*Figure 15. Performance of the model on yield prediction at plot scale.*

Among the test plots, there were plots that the network could accurately predict the yield of their sub-plots. Figure 16 illustrates an example of such a plot that the network could explain about 96 percent of yield variation among its 62 sub-plots with RMSE of 1.90 gram. In addition, Figure 16 shows a test plot that the network overestimated the yield of a substantial number of its sub-plots, and an example of a test plot that the network underestimated the yield of the majority of its sub-plots.

### 3.4.4 Yield prediction at large scale

To evaluate the feasibility of yield prediction at a large scale, the measured yield for all 50 test plots were added as a yield of a large field composed of 50 wheat plots. Alternatively, the predicted yield for these 50 plots were also added together as the predicted yield for such a large field. The total actual yield of the test plots was 59.36 kg, and the total predicted yield of these plots was 59.49 kg. Such an impressive result (i.e., about 0.2% error in yield prediction) indicates the capability of the proposed pipeline for yield prediction at a large field scale.

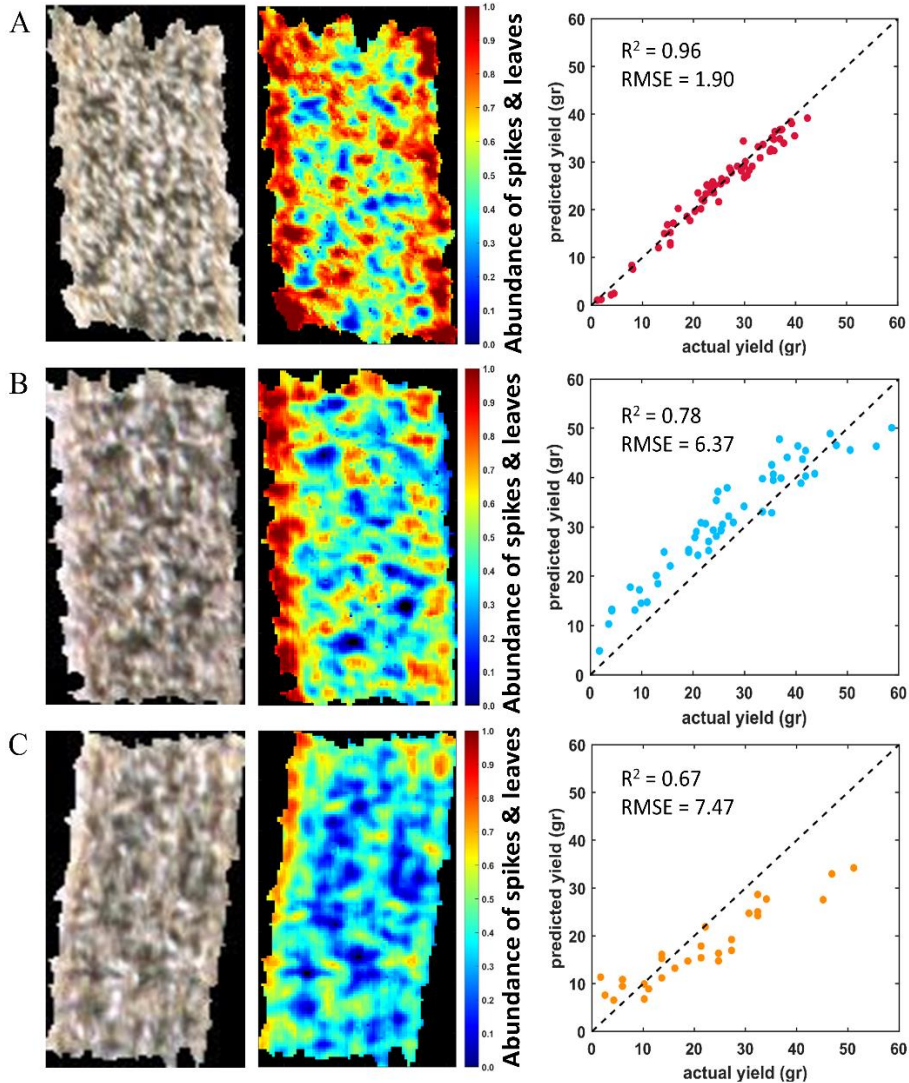

*Figure 16. Examples of test plots that the network accurately estimated (A), overestimated (B), and underestimated (C) the yield of their sub-plots. Wheat plots are wavy because the gimbal could not restrict the amplitude of vibrations caused by UAV.*

## 4 Discussion

Various methods could be used to analyze hyperspectral images for yield prediction. One of the widely used approaches is to utilize spectral indices, mostly NDVI. While yield prediction would be more accurate toward the end of growing season and prior to harvesting when the density of crop canopy is moderate to high, a method based on NDVI suffers from saturation issues at this stage of crop growth (Asrar et al., 1984; Gitelson, 2004; Gitelson et al., 1996). Therefore, this method is not suitable for yield prediction while it is extensively used by agricultural research community.

The other possible method is to train a model to directly predict the yield at plot scale. For such a model, the spectral response of pixels representing spikes and leaves in a single plot are averaged to have one feature vector since there was a single target value (i.e., measured yield) per each plot. One main drawback of this naïve approach is that substantial spectral information is suppressed by taking the spectral average over hundreds of SL pixels in a plot. Moreover, the spatial information attained with high resolution is diminished through the averaging process. The other disadvantage of taking the average across the plot pixels is that the number of samples is limited to the number of plots which was about 1000 in this study. This low number of samples might be insufficient to recognize the complex pattern from a high dimensional dataset with 190 features (i.e., number of bands) to develop a robust model for yield prediction. Lastly, investigating yield variation within plots would not be possible with this approach.

This study proposed an innovative method for analysis of high-dimensional hyperspectral images captured at high spatial and spectral resolution to estimate the yield of hundreds of wheat lines. Aerial hyperspectral images were captured in less than 10 minutes from each field using an autonomous platform. Several image processing techniques and an optimization algorithm were integrated with the domain knowledge to segment the plots from background, divide them into sub-plots, unmix the plot pixels, and assign a yield value to each sub-plot. Subsequent to these analyses, the OBIA approach was deployed to extract features from each sub-plots. Finally, deep neural networks were used to estimate the yield at sub-plot and plot scale. The results achieved by the proposed analysis framework are discussed in this section.

## 4.1 Spectral mixture analysis

With the spatial resolution of 2 cm, each pixel could potentially be a mixed-pixel, a spectral mixture of more than one particular endmember. Once the spectral signature of the endmembers was discovered from the hyperspectral image with 0.5 cm spatial resolution, the spectral mixture analysis was performed to identify the abundance of the endmembers in a given pixel. The benefits of un-mixing the pixels can be summarized into twofold. First, it allowed segmenting the plot pixels with high abundance of wheat leaves and spikes and disregarding the pixels representing background for further processing. Second, this approach provided the opportunity of deploying more advanced techniques for further investigation of yield plots by assigning a yield value to a given sub-plot based on the number of SL pixels in that sub-plot. As a result, it alleviated the curse of dimensionality by increasing the number of samples (i.e., 51,710) compared to the number of features (i.e., 381).

## 4.2 Yield analysis at sub-plot scale

Besides the yield potential of a wheat variety, the ability to produce a uniform yield across the plot is a valuable factor that could assist breeders in selecting advanced lines. However, harvesting the grains at sub-plot scale to study the yield variation within plots for various wheat varieties is not practical, particularly in a large nursery.

In this study, a novel approach was proposed to investigate the yield variation at sub-plot scale (15×15 pixel equates to 30 cm by 30 cm). First, plots were divided into sub-plots, and then a yield value was assigned to each sub-plot by integrating image processing techniques and expert domain knowledge (sub-plot yield is proportion to the number of spikes and leaves pixels representing the plant biomass). This approach offered the chance to investigate the feasibility of yield estimation at sub-plot scale with very high spatial resolution, and further evaluate the performance of various wheat lines in terms of producing yield uniformly distributed across the plot.

The results of the yield analysis at sub-plot scale revealed the significance of marginal effects on the distribution of spikes and leaves for various wheat varieties. While a particular variety might be capable of producing a uniform yield across the plot as the plants were able to compete with their neighbors, another variety might be sensitive to the plant density, causing non-uniform yield production. A uniform yield production is a fundamental trait because plants should maintain their potential yield in a competitive environment at field scale.

Figure 17 shows two wheat lines (presented in A and C) that suffer from marginal effects, and two wheat lines (presented in B and D) that produce less yield but in a more uniform manner with less marginal effects. According to the colormaps showing the distribution of the spikes and leaves in Figure 17, lines A and C produced less yield inside of the plot and more yield at the margins of the plot. Therefore, breeders prefer line B and D because of their potential in producing a more uniform yield.

This new insight about yield variation within plots casts doubts on the assumption that the performance of wheat lines in these small plots equates to performance in farmers' fields that may be up to hundreds of hectares in size. However, testing hundreds to thousands of lines for grain yield necessarily requires them to be grown in small plots (e.g. 1-10 m$^{-2}$) due to space limitations. Therefore, the proposed method is of great interest to breeders to eliminate the lines with non-uniform yield production from the breeding program because their yield performance would be significantly deteriorated in farmers' large fields where the plant density and competition is more similar to the middle rows in these small experimental plots.

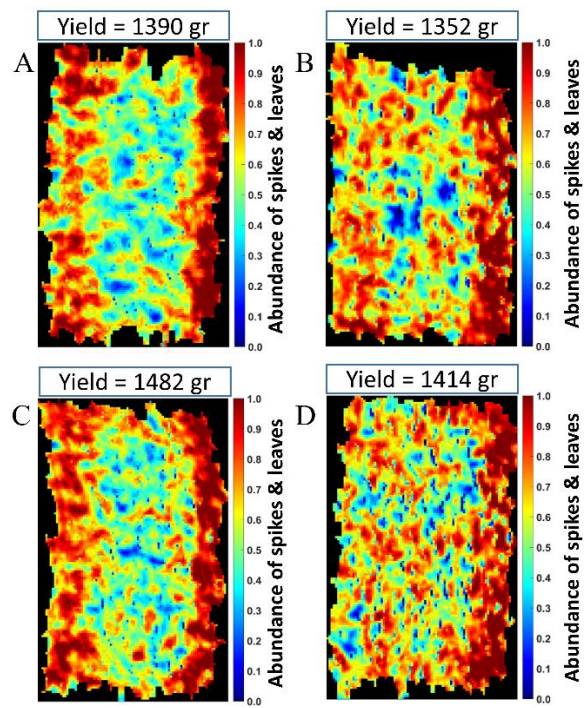

*Figure 17. (A) Example of two wheat lines that produce more yield at the margins of plots. (B) Example of two wheat lines that produce less yield but with a more uniform distribution. Despite the less yield production, breeders prefer the plots presented in (B) because of uniform yield production. Please note that wheat plots might seem wavy because the gimbal could not restrict the amplitude of vibrations caused by UAV.*

## 4.3 Practical applications of the proposed framework

The proposed framework including aerial imagery and hyperspectral image analysis can be deployed as a valuable decision support tool in breeding programs. In following sections, it is explained how this game changer framework can facilitate the process of high-throughput yield phenotyping.

### 4.3.1 Remote visual inspection of the plots

Breeders visually inspect the nursery multiple times during the growing season to record any incidents that might affect their screening, such as damages caused by animal or severe weather condition. Noticeably, this is an extremely demanding, time-consuming, and subjective task. Aerial imaging followed by the proposed automated analysis pipeline can facilitate the visual inspection to be performed remotely with high temporal resolution and across all nurseries in multiple locations. Figure 18 shows the SL colormap obtained by analysis of aerial hyperspectral images conformed to the notes taken by an expert in the field. The existence of more SS pixels (presented in blue color) implies less yield regardless of the variety because it indicates the pixels representing soil and shadow, which do not contribute to the yield.

The presented framework can serve as a tool to remotely inspect the status of plots and accordingly make an appropriate decision. For instance, based on the SL colormap, a breeder would disregard the plot shown in Figure 18A because the measured yield value for the corresponding wheat line is not a reliable indicator due to the severe damage. In addition, this method can assist breeders in identifying low-yielding plots prior to harvesting.

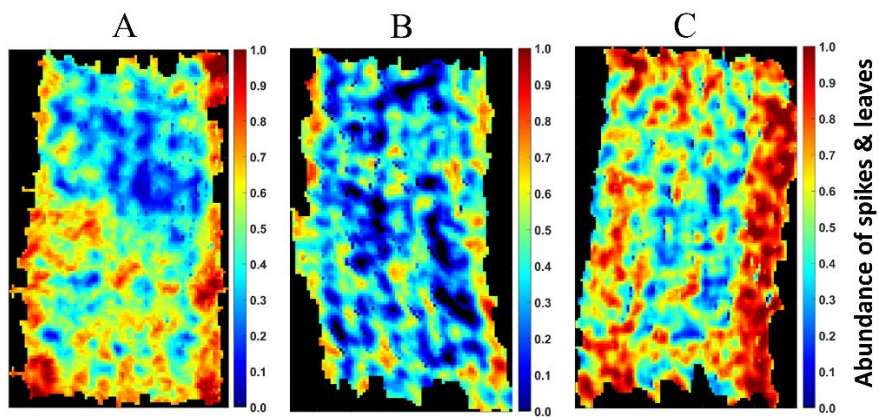

*Figure 18. Remote visual inspection of the plots by analyzing the aerial hyperspectral images. The results of the image analysis conformed with the notes made by an expert in the field. The notes made for these plots were: (A) ½ of the plot was damaged, (B) weak plot, (C) strong plot. Please note that wheat plots might seem wavy because the gimbal could not restrict the amplitude of vibrations caused by UAV.*

### 4.3.2 Investigating more lines by optimizing plot size

Currently, several factors dictate the plot size for yield trials, including seed availability (primarily a concern for 1st year yield trials only), cost and availability of land, size of available small plot equipment, minimizing experimental error, and overall cost of labor and resources. All of these factors become even more restricting since a breeder often manages yield trials in multiple locations to account for non-uniform climate, soil, and environmental conditions. Therefore, these factors restrain the number of 1st year yield plots that a breeding program can manage, and subsequently, they affect the size of the succeeding yield trials.

The ability of aerial imagery in remote inspection and yield prediction of the plots can reduce the required labor and equipment for scouting and harvesting. In addition, the unique advantage of the proposed framework in yield estimation with high spatial resolution enables plant scientists and breeders to optimize the plot size and investigate more wheat lines in a dedicated field each year. During the first few years, wheat lines can be planted in smaller plots, and the proposed framework can be utilized to perform a fast binary screening based on their yield performance (Figure 19). Low-yielding lines will be discarded, and only high-yielding plots are harvested to obtain seeds for the next trial. This allows breeders to manage their labor, equipment, and land in a more effective manner. Afterward, the high-yielding lines are evaluated with larger plot size in more locations over the following years. While number of lines decreases over time, the size of the plot increases to evaluate the yield performance of advanced lines in environments more similar to growers' field conditions.

### 4.3.3 Other potential applications

In addition to yield estimation, breeders can utilize the proposed framework to: (i) study the effect of plant density on yield with high spatial resolution, (ii) study the impact of side trimming on yield across various varieties, and (iii) investigate multiple desired traits, such as disease resistance, at the early stages of selecting advanced wheat lines.

## Author Contributions

AM conducted the aerial hyperspectral imagery, performed the preprocessing of images, and developed the entire framework for analysis of images. CY provided technical assistance in analysis of images. JA was responsible for breeding of wheat lines, managing the nurseries, and collecting the ground-truth data. AM wrote the paper, and CY and JA edited the paper.

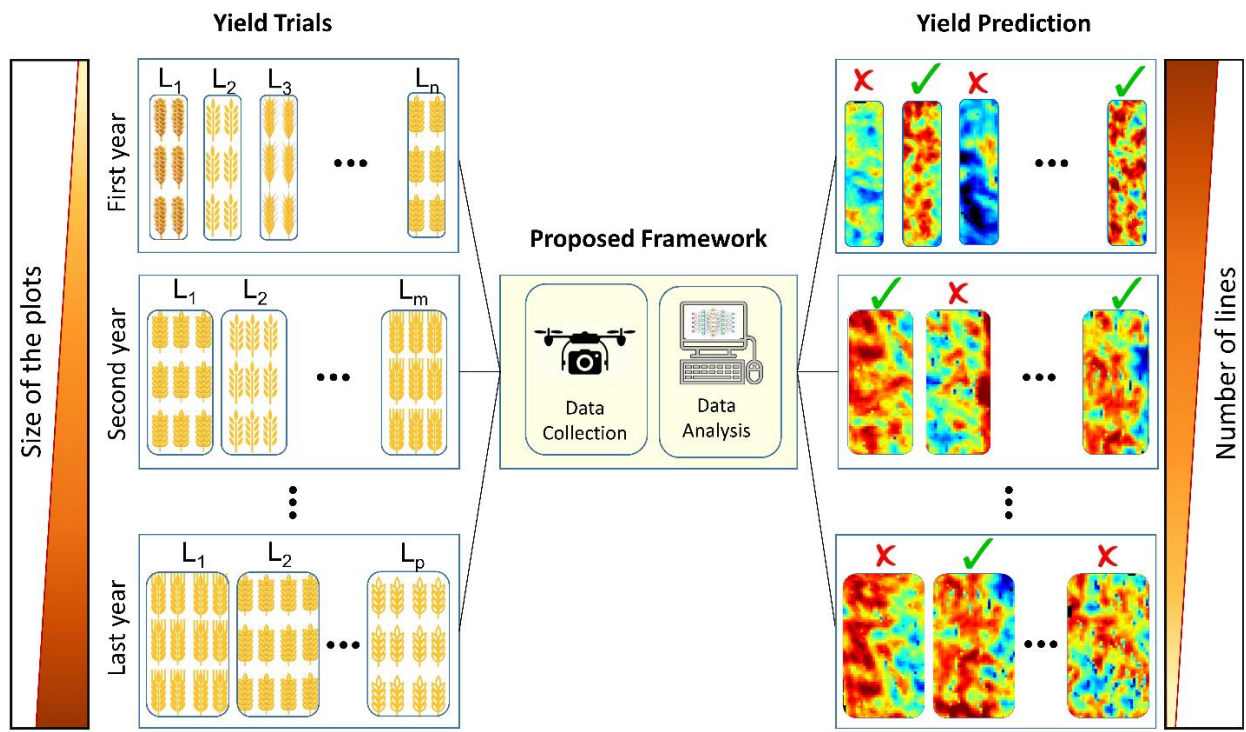

*Figure 19. Nominating advanced lines for commercialization over several years of yield trials using the proposed framework. In each year, aerial hyperspectral imagery followed by the proposed analysis pipeline is utilized to classify the wheat lines into low- and high-yielding lines based on their yield performance. While low-yielding lines are discarded, high-yielding lines are advanced to the next years' yield trial with larger plot size to evaluate yield performance in environments more similar to grower's field conditions.*

## Acknowledgment

The authors would like to gratefully acknowledge the funding from the Minnesota's Discovery, Research, and InnoVation Economy (MnDRIVE) program through the research area of Robotics, Sensors, and Advanced Manufacturing. We thank Ms. Susan K. Reynolds for her valuable support in managing the fields and collecting the ground truth data, and Mrs. Parisa Kafash for her assistance in preparing the figures. We would also like to acknowledge the graduate student fellowships provided by MnDRIVE Global Food Ventures and the department of Bioproducts and Biosystems Engineering.